\newcommand{\diracslash}[1]{#1\llap{/\kern2pt}}

\newcommand{\be}{\begin{equation}}
\newcommand{\ee}{\end{equation}}
\newcommand{\bea}{\begin{eqnarray}}
\newcommand{\eea}{\end{eqnarray}}
\newcommand{\ba}[1]{\begin{array}{#1}}
\newcommand{\ea}{\end{array}}

\documentclass[prd,aps,floats,nofootinbib,tightenlines,showpacs]{revtex4}
\usepackage{epsfig,graphicx,pstricks}
\usepackage{psfrag}
\usepackage{color}
\usepackage{amsmath}
\usepackage{amsfonts}
\usepackage{amssymb}
\usepackage{color, soul}
\usepackage{textcomp}
\usepackage{multirow}
\usepackage{natbib}
\usepackage{subfigure}
\begin{document}
\title{Study of net-baryon higher moments in PNJL model and their expectation for net-proton using the Subensemble Acceptance Method for the search of QCD critical point}
\author{A.~Sarkar}
\email{amal.sarkar@cern.ch}
\author{P.~Deb}
\email{paramita.dab83@gmail.com}
\author{Bidhan Mandal}
\email{mandal.bidhan1440@gmail.com}
\author{R.~Varma}
\email{raghava.varma@cern.ch}
\affiliation{*School of Physical Science, Indian Institute of Technology Mandi, Kamand, Mandi - 175005, India}
\affiliation{{\dag\S}Department of Physics, Indian Institute of Technology Bombay, Powai, Mumbai- 400076, India}
\affiliation{{\ddag}Department of Physics, Ramakrishna Mission Residential College (Autonomous), Narendrapur, Kolkata - 700103, India}

\begin{abstract}
One of the most important parts of the QCD phase diagram of strongly interacting matter is the Critical End Point. The non-monotonic behavior of the conserved quantities like net-baryon ($\Delta B$), net-charge ($\Delta Q$), and net-strangeness ($\Delta S$) are believed to be the signatures of the QCD Critical End Point (CEP) as a function of the energy. We study the effect of the QCD critical point on moments of net-baryon in the Polyakov loop enhanced Nambu-Jona-Lasinio (PNJL) model of QCD with six quark and eight quark interactions. The study is performed at energies similar to RHIC beam energy scan (BES). Experimentally measuring conserved quantities is difficult due to systematic limitations, therefore net-proton, net-pion, and net-kaon are measured as the proxy of $\Delta B$, $\Delta Q$, and $\Delta S$. Thus the need for different models becomes predominant to estimate the value of different observables. Higher-order moments like skewness ($S$), kurtosis ($\kappa$), and their system volume independent products ($ M/\sigma^{2}, s\sigma$, $\kappa\sigma^{2}$) which are calculated in the PNJL model, are sensitive to the produced correlation length of the hot and dense medium, making them more prone to search for the critical point. Recent studies in the subensemble acceptance method (SAM) on the HRG model shows the dependency of the measure higher order moment on the experimental acceptance. We used SAM to analyze the behavior of $\kappa\sigma^{2}$ of net baryon distribution within the subvolume system for various acceptance fractions. These results can be directly mapped to the percentage of the subvolume (particle) of the total volume (conserved quantities). The results are compared to the STAR net-proton and proton data with different energies to understand the existence of critical point. For reference, results are also compared with the theoretical UrQMD and HRG models.
\end{abstract}
\maketitle
\section{\label{sec:level1}Introduction
}
Quantum Chromodynamics (QCD) describes the strong interaction based on $SU(3)$ gauge theory with quark in the fundamental and gluon in the adjoint replacement of the gauge group weak connection. Hadrons confine quarks and gluons at ordinary temperatures but a de-confinement phase of quarks and gluons takes place at very high temperatures and density, creating a state called Quark-Gluon Plasma (QGP) \cite{1}. It is observed that some matter undergoes a change from a hadronic state to a QGP state initiating a first-order chiral phase transition. A rapid but smooth cross-over transition at a vanishing chemical potential ($\mu_{_B}$) and large temperature T is predicted by lattice QCD calculation, whereas different models modeling matter at vanishing T anticipate a strong first-order phase transition at a large $\mu_{_B}$. If various models are correct, then a critical point must be there at finite $\mu_{_B}$, where the characteristics of the transition change from a smooth cross-over to first-order \cite{2, 3}. The critical endpoint at the end of the first-order transition line is a crucial point where the phase transition is second-order \cite{4, 5}. Numerous effective models have been used to do extensive research on the characteristics of low-energy hadrons and the nature of the chiral phase transition at infinite temperature and density \cite{6, 7, 8}. The Nambu-Jona-Lasinio (NJL) model \cite{9} describes the interaction of constituent quark fields. In place of the local gauge $SU(3)$ color transformation of the QCD Lagrangian, it presents a global $SU(3)$ symmetry. The NJL dynamics reduce this color confinement \cite{10}. Specifically, the induction of a charged pion condensation phenomenon has been used to represent the limited size effect of a dense baryonic matter in a $1 + 1$ dimensional NJL model. An important development in this exploration was the proposal for expanding the chiral Lagrangian, such as the NJL model, by connecting quarks to a uniform temporal background of the gauge field (the Polyakov loop). The PNJl model \cite{11} demonstrates two essential characteristics of QCD, which are spontaneous chiral symmetry breaking and confinement-like properties. 
This work is of prime importance in high-energy physics and is 
not confined to theoretical approaches only but also strived towards exploring CEP in the experimental domain. One such approach is the heavy-ion collider experiments where we can collide ultra relativistic particles to get the elementary constituents of matter. But the chief constraint lies in the fact that the elementary constituents produced by the instantaneous collision of particles are evanescent in nature and thus very difficult to observe directly. The event-by-event fluctuations or higher order moments of conserved quantities such as net-charge ($\Delta{N_{ch}}$), net-baryon ($\Delta{N_{B}}$), and net-strangeness ($\Delta{N_{S}}$) number are predicted to depend on the produced non-equilibrium correlation length $\xi$ \cite{12, 13} in the system. In an idealized thermodynamic limit, the correlation length of the system which is related to the magnitude of the non-Gaussian fluctuations of the conserved quantities diverges at the critical point. Theoretical calculation suggests that due to the long-range correlations at all length scales $\xi$ may rise from $\sim$ 1.5 to 3 fm in the heavy-ion collision, depending on the system size. Due to the experimental limitation measuring net baryons is very difficult therefore, protons and anti-protons are measured and have been shown to be reliable proxies for baryons and anti-baryons. Therefore, the study on experimental acceptance is very important to understand the results. Also, a comprehensive study is required to rely on the net-proton results as a proxy for net-baryon. The subensemble acceptance method is such an effective method, which can give us a clear idea of the confidence level on the results of experimentally measured proxies \cite{40}. In heavy-ion collision experiments, the QCD critical point can be found via the non-monotonic behavior of many fluctuations observable as a function of the collision energy. It is necessary to scan the phase diagram for the search of the critical point. In experiment it can be done by adjusting the initial collision energy $\sqrt{s}$ while changing the temperature and chemical potential. The moments of the event-by-event multiplicity distributions can be used to describe the fluctuations in the event-by-event particle multiplicity. The magnitude of fluctuations in conserved quantities at finite temperatures is distinctly different in the hadronic and QGP phase. Higher-order moments like mean (M), variance ($\sigma^2$), skewness (S), kurtosis ($\kappa$) of a conserved quantities depend on the higher power of $\xi$ i.e., $ \sigma\sim\xi^{2}, S\sim\xi^{4.5}$ and $\kappa\sim\xi^{7}$ \cite{16}. In the case of freeze-out going through a phase transition or critical point and the signal survives the system's evolution, it can be identified by measuring the non-monotonic behavior of the measured moments. The significance of the intended signals is constrained by the finite size and time effects in heavy-ion collisions \cite{14}. The Beam Energy Scan (BES) at the Relativistic Heavy-Ion Collider (RHIC) is designed to investigate bulk QCD features such as phase transition, QCD critical point, and thermalization of matter. The tremendous quantity of energy deposited in a very tiny region of space for a very short period of time is the most significant property of relativistic nucleus-nucleus collisions. The moment ratios of experimental observables cancel the volume dependency of the system and can be directly compared to the ratios of susceptibilities from the theoretical calculation. Therefore, it might be helpful to study the higher order moments like [skewness $S={\langle(\delta N)^{3}\rangle}/{\sigma^{3}}$ and kurtosis $\kappa=[{\langle(\delta N)^{4}\rangle}/{\sigma^{4}}]-3$ where $\delta N=N-\langle N\rangle$] and their volume independent moment products of the conserved quantities as they have stronger dependence on $\xi$ \cite{15, 16}.  The critical point would therefore be strongly corroborated by the non-monotonic appearance and then the disappearance of the critical fluctuation.

The subensemble acceptance method is an important way to a direct comparison between theoretical calculations of grand-canonical susceptibilities and higher order moment of conserved charges measured in central heavy ion collisions at the highest energies where we have a strong space-momentum correlation. It quantifies the effect of global conservation laws. For a small selected acceptance window, the possible dynamical correlations will also be strongly reduced \cite{41} and consequently, net baryons will be distributed according to the difference of two
independent Poisson distributions, the Skellam distribution. On the other hand, correlations due to conservation of baryon number will become relevant by expanding the acceptance to capture dynamical fluctuations. To get a quantitative estimate of what means large acceptance, the acceptance factor for baryons is defined as the ratio of the mean number of detected baryons $\langle N^{acc}_{B}\rangle$ to the number of baryons in full phase space $\langle N^{4\pi}_{B}\rangle$. Experiments typically report on net-proton higher-order moments which are used as a proxy for the net-baryons. The acceptance fraction $\alpha$ of the net proton distribution for baryon number conservation in a sub-volume can be defined as $\alpha = \langle N^{acc}_{p}\rangle / \langle N^{4\pi}_{B}\rangle$, where the mean number of protons inside the acceptance is denoted by $\langle N^{acc}_{p}\rangle$. The Ratio of fourth-order to second-order moment $\kappa\sigma^{2} (\frac{C_{4}}{C_{2}})$ for distribution of charge $B_{1}$ within the subsystem of the total net baryon number, $B$, is defined by the suggested equation,

    \begin{equation}\label{27}
\frac{C_{4}[B_{1}]}{C_{2}[B_{1}]}=(1-3\alpha\beta)\frac{\chi_{4}^{B}}{\chi_{2}^{B}}-3\alpha\beta\\
(\frac{\chi_{3}^{B}}{\chi_{2}^{B}})^{2}
\end{equation}
where $\beta=1-\alpha$.\\




In this paper, we study the fluctuations of higher moments (mean ($M$), variance ($\sigma^2$), skewness ($S$), kurtosis ($\kappa$)) and their volume independent moment products ($ s\sigma$, $\kappa\sigma^{2}$) of net-baryon in the 3 flavors finite volume finite density PNJL model. We have elaborated the formalism of the PNJL model briefly in section III. The higher order moments and their volume independent moment products of the conserved quantities are expected to show fluctuations as they are significantly more sensitive to the proximity of the QCD phase diagram. In this work, we have also presented the results from subensemble acceptance method, which uses acceptance fraction alpha ($\alpha$) = 0.5, 0.6, 0.7, 0.8, 0.9, and 1 to evaluate the higher order moment ($\kappa\sigma^{2}$) of the charge distributions within the subsystem. The results have been compared to the STAR experimental data of net-proton and proton \cite{35} along with HRG \cite{19, 37}, the UrQMD model \cite{36} and Lattice QCD calculations \cite{39}. The nature of the data points are analyzed and compared with theoretical and experimental data. The fluctuations as a function of energy give an estimation for the critical region. 

\maketitle
\section{\label{sec:level2}Higher-order moments and volume independent moment products
}
The physics of heavy-ion collisions is one of the several experimental windows of QCD phenomenology. There are many experimental programs like the Beam Energy Scan (BES) program at the Relativistic Heavy-Ion Collider (RHIC) which aims to study the detailed QCD phase structure \cite{17, 18, 19, 20}. As any system in thermal and chemical equilibrium is characterized by the given values of temperature $T$ and baryon chemical potential $\mu_B$. Modern physicists map the phase diagram of strongly interacting matter (describe through QCD) through temperature ($T$) versus baryonic chemical potential ($\mu_B$) curve. With many studies, the existence of QCD critical point and the first-order phase boundary between quark-gluon and hadronic phases. The critical point is of great importance to modern physicists. As known, a phase transition is actually a thermodynamic singularity of the system. But as we decrease $\mu_B$ after the critical point, the transitions are smooth i.e. there is no singularity \cite{21}. Theoretically, the determination of CP at a coordinate ($\mu_B$, $T$) is a definite task that involves analytical skill in the mathematical method that is yet to be achieved at present. The only choice left is to approach it using numerical methods. This is achieved through Monte Carlo simulations. But, they also have limitations at the non-zero value of $\mu_B$. The simulation models are made to satisfy the vacuum phenomenology i.e., $T = \mu_B = 0$ \cite{22}. Therefore, the next plan of action toward the determination of CP is from experimental means. Even though the exact coordinate of CP is yet to be calculated theoretically, it is suggested from different studies that the location of CP is within a regime that can be probed through heavy-ion collision experiments. The underlying physics for the detection of signatures of CP in the experiment is interpreted as the divergence of susceptibilities at the critical points. Some observables which are related to the susceptibilities are needed which would show some fluctuation at CP and cross-over. The moments of the distributions of conserved quantities are considered to be good observable for finding the signatures of phase transition and CP. Therefore, the calculation of higher moments is necessary for the study of the QCD phase. Theoretical and experimental challenges faced during the determination of CP are described in Ref. \cite{23}.\\

\subsection*{\textbf{A. Higher-order moments and cumulants}}
In terms of statistics, a probability distribution is mainly characterized by various higher-order moments (Mean (M), variance ($\sigma^2$), Skewness (S), and kurtosis ($\kappa$) ). This section is dedicated to the definition of central moments and cumulants. The distribution of the conserved quantity can be simply given by the major emitted particle of the quantity. For example, net-proton distribution is used as a proxy for net-baryon quantity as the abundance of the proton is most among other emitted baryons (about 70$\%$). Similarly, net-kaon distribution is used as a proxy of net-strangeness quantity $\sim$ 65$\%$. Experimentally, Net-proton distribution is find out by calculating event-by-event $\Delta{N_{p}}$ as,
\begin{equation}\label{1}
\Delta N_P=N_P-N_{\bar{P}}
\end{equation}
	
where $N_P$ is the proton number and $N_{\bar{P}}$ is the anti-proton number produced in the collision. Generally, $\Delta N$ is used to denote the Net-particle number (in the case of proton $\Delta N_P$) in one event. Thus, $\left\langle N\right\rangle $ represents the average of N over all event ensemble. Here, angular brackets are used to denote the average of the event-wise distribution. And the deviation of N from its average net-particle multiplicity (\emph{i.e.} M=$\left\langle N\right\rangle $) is given by,
	\begin{equation}\label{2}
	\delta N = N-M
	\end{equation}
	While defining the central moment, the average of net-particle distribution is denoted as,
	\begin{equation}\label{3}
	\hat{\mu}=\left\langle N\right\rangle =M
	\end{equation}
	The hat is used to define the average operation. The $r^{th}$ central moment is given by,
	\begin{equation}\label{4}
	\hat{\mu_{r}}=\left\langle \left( \delta N\right)^r \right\rangle	
	\end{equation}
	Also,
	\begin{equation}\label{5}
	\hat{\mu_{1}}=\left\langle N\right\rangle-\hat{\mu}=\hat{\mu}-\hat{\mu}=0
	\end{equation}
	The cumulants can be calculated from the moments as,
	\begin{equation}\label{6}
    \begin{split}
	\hat{C_1}=\hat{\mu} \\
	\hat{C_2}=\hat{\mu_2} \\
	\hat{C_3}=\hat{\mu_3}
    \end{split}
	\end{equation}
	Above n=3, $\hat{C_n}$ is generalized through a recursion relation as [24],
	\begin{equation}\label{7}
	\hat{C_n}=\hat{\mu_n}-\sum_{m=2}^{n-2}\binom{n-1}{m-1}\hat{C_m}\hat{\mu_{\left( n-m\right) }}
	\end{equation}
	Thus, the skewness and kurtosis of the distribution is denoted as,
	\begin{equation}\label{8}
	\hat{M}=\hat{C_1}\text{ ; }\hat{\sigma^2}=\hat{C_2}\text{ ; }\hat{S}=\frac{\hat{C_3}}{\hat{C_2}^{3/2}}\text{ ; }\hat{\kappa}=\frac{\hat{C_4}}{\hat{C_2}^{2}}
    \end{equation}
The moments define the characteristics of the net-particle multiplicity distribution. The first-order moment describes an expectation operator of the multiplicity density. The second-order moment is called variance and it gives the susceptibility of the measurements. The third-order moment measures the lopsidedness or nature of the tail of the distribution. The normalized third-order moment is known as skewness which provides information on the direction of variation of the distribution. The fourth-order moment compares the peakness or shortness and squatness of a distribution. The normalized fourth-order moment is known as heteroskedasticity or kurtosis. The kurtosis actually comes from the normalized fourth-order moment minus 3. The subtraction of 3, originates from the Gaussian distribution. The fifth-order moment measures the asymmetry sensitivity of the fourth-order moment and in general, compound options are related to the sixth-order moment. The detailed calculation of moments from the statistical point of view from the definition of the Partition function is calculated analytically \cite{24}. The non-linear sigma model (NLSM) is a phenomenological method that enables the study of critical opalescence in nuclear systems. In this model, the moments are also related \cite{25, 26} to the correlation length of the system produce in the heavy-ion collision, given by $\xi$. From, theoretical calculation it is found that for heavy-ion collision, $\xi \approx 1.5-3 \text{ fm}$. The variation of correlation length to the moments are given as,
\begin{equation}\label{9}
\hat{\sigma^2} \sim \xi^2 ;\\
\hat{S} \sim \xi^{4.5} ;\\
\hat{\kappa} \sim \xi^{7}
\end{equation}
The variation of the correlation length to the moments is given in Ref. [16],\\
\subsection*{\textbf{B. Product of moments and system volume dependency}}
The thermodynamic susceptibilities, $\chi_i^{\left( n\right) }$, where \emph{n} is the order of the susceptibility and \emph{i} stands for the type of conserved quantum number are related to the correlation lengths of the system which are again related to the higher order moments of conserved quantities. Therefore, the moments of conserved quantities are related to these susceptibilities. For net-proton distribution, \emph{i} becomes baryon quantum number (\emph{B}). These susceptibilities are written in terms of cumulants ($C_n$) as,
	 \begin{equation}\label{10}
	 \chi_i^{\left( n\right) } = \frac{1}{VT^3}C_n
	 \end{equation}	
	 where V is the system volume and T is the temperature.
	
From Eq. (11) susceptibilities are dependent on volume i.e. the size of the system. This can also be observed that most of the cumulant values show a linear variation with the average number of participants (\emph{viz.} $\left\langle N_{part}\right\rangle $) which is calculated from the globular model, this means a linear increase with $\left\langle N_{part}\right\rangle $) as the system volume increases. Refer to \cite{19, 20} for net-proton, net-kaon, and net-$\Lambda$ particle multiplicity distribution respectively. For volume independent susceptibility ratios, moment products or ratios of cumulants of net-particle multiplicity distributions are calculated. They are related as,
	 \begin{equation}\label{11}
	 	\frac{\sigma^2}{M}=\frac{\chi_i^{\left( 2\right) }}{\chi_i^{\left( 1\right) }};\\
	 	S\sigma=\frac{\chi_i^{\left( 3\right) }}{\chi_i^{\left( 2\right) }};\\
	 	\kappa\sigma^2=\frac{\chi_i^{\left( 4\right) }}{\chi_i^{\left( 2\right) }}
	 \end{equation}
The calculation of the product of moments is easily done by using Eq. (8) as follows,
	 \begin{equation}\label{12}
	 \frac{\sigma^2}{M}=\frac{C_2}{C_1}\text{ ; }S\sigma=\frac{C_3}{C_2}\text{ ; }\kappa\sigma^2=\frac{C_4}{C_2}\text{ ; }\frac{\kappa\sigma}{S}=\frac{C_4}{C_3}
	 \end{equation}\\

\maketitle
\section{\label{sec:level3} The Polyakov loop enhanced Nambu-Jona-Lasinio model}
At low temperatures, lattice and phenomenology say the state of the matter is confined. Whereas at high temperatures, perturbation theory is reliable and shows de-confinement. Therefore, there is a need to build a phenomenological model that incorporates both low and high temperature QCD behaviour in a single picture. In PNJL model the interactions between quarks with the temporal gluon field is represented by the Polyakov loop. PNJL is an extension of the Nambu-Jona-Lasinio model, that can serve this purpose. According to this model, the chiral point couplings between quarks and a background field that symbolises Polyakov loop dynamics serve as a framework for gluon dynamics. In this model, instead of point-like four-fermion coupling a non-local interaction is employed. Due to the symmetry of the Lagrangian in PNJL model makes it the same versatile class as expected for QCD. Therefore, for the test of critical phenomena related to the breaking of chiral symmetry and global Z(3) this model can play an important role. 
In the pure gauge system, the first-order de-confinement phase transition \cite{30} becomes discontinuous in the Polyakov loop, which becomes a smooth crossover when quarks are introduced. A representation of the Polyakov line is \cite{31}

\begin{equation}\label{13}
L(\vec{x})={p}\exp[i\int^{\beta}_{_0}d\tau A_{_4}(\vec{x},\tau)]
\end{equation}
where $A_{_4} = iA_{_0}$, is the temporal component of Euclidean gauge field $(\vec{A},A_{_4})$, which has absorbed the strong coupling constant  $g_{_s}$. $p$ signifies path ordering. $\beta=1/T$ using the Boltzmann constant $K_{_B}=1$. Under global $Z(3)$ symmetry, the Polyakov line $L(\vec{x})$ evolves to a field with charge one.
The Polyakov loop field \cite{32} is represented as
\begin{equation}\label{14}
\begin{split}
    \Phi=(Tr_{_c}L)/N_{_c} \\
    \text{and\quad it\'s\quad conjugate\quad} \bar{\Phi}=(Tr_{_c}L^{\dag})/N_{_c}
  \end{split}
\end{equation}
The NJL model describes the interaction of the constituent quark fields, with the exception of the replacement of a covariant derivative including the temporal background gauge field $\Phi$.
Thus, the 2+1 flavour version of the PNJL model is explained using the Lagrangian \cite{33},
\begin{equation}\label{15}
\begin{split}
  \pounds=\sum_{_{f=u,d,s}}\bar{\psi}_{_f}\gamma_{_\mu}iD^{\mu}\psi_{_f}-\sum_{_f} m_{_f}\bar{\psi}_{_f}\psi_{_f}
  +\sum_{_f}\mu_{_f}\gamma_{_0}\bar{\psi}_{_f}\psi_{_f}+\\
  \frac{g_{_s}}{2}\sum_{_{a=0,...,8}}[(\bar{\psi}\lambda^{a}\psi)^{2}
  + (\bar{\psi}i\gamma_{_5}\lambda^{a}\psi)^{2}]-g_{_D}[det \bar{\psi}_{_f}P_{_L}\psi_{_{f^{\prime}}}\\
  + det \bar{\psi}_{_f}P_{_R}\psi_{_{f^{\prime}}}]-u^{\prime}[\Phi,\bar{\Phi},T]
\end{split}
\end{equation}
where the flavors indicated by $f$ are, respectively, u, d, and s. For two flavors $g_{_D}=0$. $m_{_f}$ represents the diagonal elements of mass matrix $(m_{_u},m_{_d},m_{_s})$. The flavour $SU_{_f}(3)$ Gell-Mann matrices $(a=0,1,..,8)$ are also represented by $\lambda^{a}$, where $\lambda^{0}=\surd(\frac{2}{3})I$.
The left-handed and right-handed chiral projectors, denoted by the matrices  $P_{_{L, R}}=(1\pm\gamma_{_5})/2$, respectively \cite{34}. $D^{\mu}=\partial^{\mu}-iA^{\mu}$, with $A^{\mu}=\delta^{\mu}_{_0}A_{_0}$ (Polyakov gauge); in Euclidean notation $A_{_0}=-iA_{_4}$ is used to indicate the covariant derivative.
The Polyakov loop, which is the normalized trace of the Wilson line $L$, ought to exceed unity in the PNJL model above $2T_{c}$. This issue can be resolved by applying a suitable Jacobian transformation from the matrix-valued field $L$ to $\phi$, which will then limit the value of $\phi$ to a value inside 1.
So, for reproducing lattice results, the Vandermonde term is to be introduced in the Polyakov loop potential. Therefore the potential $u^{\prime}$ with the vandermonde term can be written as
\begin{equation}\label{16}
 \frac{u^{\prime}(\Phi,\bar{\Phi},T)}{T^{4}}=\frac{u(\Phi,\bar{\Phi},T)}{T^{4}}-\kappa ln[ \jmath(\Phi,\bar{\Phi})]
\end{equation}
where $u(\Phi)$ is the Landau-Ginzburg type potential equivalent to $Z(3)$ global symmetry. we choose [32],
\begin{equation}\label{17}
 \frac{u(\Phi,\bar{\Phi},T)}{T^{4}}=-\frac{b_{_2}(T)}{2}\bar{\Phi}\Phi-\frac{b_{_3}}{6}(\Phi^{3}+\bar{\Phi}^{3})
 +\frac{b_{_4}}{4}(\bar{\Phi}\Phi)^{2}
\end{equation}
with $ b_{_2}(T)=a_{_0}+a_{_1}(\frac{T_{_0}}{T})+a_{_2}(\frac{T_{_0}}{T})^{2}+a_{_3}(\frac{T_{_0}}{T})^{3}$ and $T_{_0}$ is the critical temperature for de-confinement phase transitions according to pure gauge lattice theory. $b_{_3}$ and $b_{_4}$ are being constant.
The vandermonde term $\frac{u(\Phi,\bar{\Phi},T)}{T^{4}}$ duplicates the effect of $SU(3)$, where $\jmath(\Phi,\bar{\Phi})$ is the Jacobian of transformation from the Wilson line $L$ to $(\Phi,\bar{\Phi})$ written as
\begin{equation}\label{18}
  \jmath[\Phi,\bar{\Phi}]=[(\frac{27}{24\pi^{2}})(1-6\Phi\bar{\Phi}+4(\Phi^{3}+\bar{\Phi}^{3})-3(\Phi\bar{\Phi})^{2}]
\end{equation}
$\jmath[\Phi,\bar{\Phi}]$ is also known as the vandermonde determinant. This determinant does not explicitly depend on space-time. The value of $\kappa$ will be determined phenomenologically. In LQCD computation these parameters are obtained by fitting a few physical variables as a function of temperature \cite{35}.  The set of selected values are listed in the table \ref{table1} [30].\\
\begin{table}[htb]
\begin{center}
\begin{tabular}{|c|c|c|c|c|c|c|c|c|c|c|c|}
\hline
Interaction & $ T_0 (MeV) $ & $ a_0 $ & $ a_1 $ & $ a_2 $ & $ b_3 $ &$
b_4$ & $  \kappa $ \\ 
\hline
6-quark &$ 175 $&$ 6.75 $&$ -9.0 $&$ 0.25 $&$ 0.805 $&$7.555 $&$ 0.1 $ \\
\hline
8-quark & $ 175 $&$ 6.75 $&$ -9.8 $&$ 0.26 $&$0.805$&$ 7.555 $&$ 0.1 $\\
\hline
\end{tabular}
\caption{Parameters for the Polyakov loop potential of the model.}  
\label{table1}
\end{center}
\end{table}
\\\vspace{-0.5cm}

The NJL part of the theory has similarities to the BCS theory of superconductor, where the pairing of two electrons leads to condensation causing a gap in the energy spectrum. 
Similar to this, in the chiral limit, the NJL model displays dynamical breaking of $SU(3)_{_L}\times SU(3)_{_R}$ symmetry to $SU(3)_{_V}$. As a result, a composite operator detects a nonzero vacuum expectation value leading to $\langle\bar{\psi}_{_f}\psi_{_f}\rangle$ condensation. The quark condensate is given by
\begin{equation}\label{19}
 \langle\bar{\psi}_{_f}\psi_{_f}\rangle=-i N_{_c}\pounds t_{_{y\rightarrow x^{+}}}(trS_{_f}(x-y)),
\end{equation}
where the trace is over the states of color and spin. The self-consistent gap equation for the constituent quark masses are
\begin{equation}\label{20}
  M_{_f}=m_{_f}-g_{_s}\sigma_{_f}+g_{_D}\sigma_{_{f+1}}\sigma_{_{f+2}}
\end{equation}
where the chiral condensate of the quark with flavor $f$ is denoted by  $\sigma_{_f}=\langle\bar{\psi}_{_f}\psi_{_f}\rangle$. Here if we consider $\sigma_{_f}=\sigma_{_u}$, then $\sigma_{_{f+1}}=\sigma_{_d}$ and $\sigma_{_{f+2}}=\sigma_{_s}$. Similarly if $\sigma_{_f}=\sigma_{_d}$, then $\sigma_{_{f+1}}=\sigma_{_s}$ and $\sigma_{_{f+2}}=\sigma_{_u}$; if $\sigma_{_f}=\sigma_{_s}$, then $\sigma_{_{f+1}}=\sigma_{_u}$ and $\sigma_{_{f+2}}=\sigma_{_d}$. At zero chemical potential and temperature \cite{36}, the expression for $\sigma_{_f}$ can be expressed as
\begin{equation}\label{21}
\sigma_{_f}=-\frac{3M_{_f}}{\pi^{2}}\int^{\Lambda}_{_0}\frac{p^{2}}{\sqrt{(p^{2}+M^{2}_{_f})}}dp
\end{equation}
where $\Lambda$ is the three-momentum cutoff. Because of the dynamical breaking of chiral symmetry, $N^{2}_{_f}-1$ Goldstone bosons appear \cite{37}. The masses and decay width from experimental observations of these Goldstone bosons, which are pions and kaons are utilized to further determines the parameters of NJL model. In the present study, two sets of parameters have been considered for the quark sector: one with six quark interactions where g1 and g2 are zero and another with up to eight quark interactions where g1 and g2 have finite values.  Moreover, it has been observed that the interaction of four quark forms a stable vacuum in the quark sector by breaking the chiral symmetry spontaneously. However, the six-quark interaction term should reproduced the UA(1) anomaly to destabilize the vacuum. Further increasing the quark interaction up to eight quarks stabilizes the vacuum. In order to consider the stability of the PNJL model and to determine if the fluctuations change significantly from six to eight quark interactions, two distinct parameter sets are required. The set of parameters is tabulated in table \ref{table2} for six and eight quark interactions.

\begin{table}[htb]
\begin{center}
\begin{tabular}{|c|c|c|c|c|c|c|c|c|c|c|}
\hline
Model & $ m_u (MeV) $ & $ m_s (MeV)$ & $ \Lambda (MeV) $ & $ g_S \Lambda^2 $ & $ g_D \Lambda^5 $ 
&$
g_1 \times 10^{-21} (MeV^{-8})$ & $ g_2 \times 10^{-22} (MeV^{-8})$ \\ 
\hline
With 6-quark &$ 5.5 $&$ 134.76 $&$ 631 $&$ 3.67 $&$ 9.33 $&$0.0 $&$ 0.0 $ \\
\hline
With 8-quark & $ 5.5 $&$ 183.468 $&$ 637.720 $&$ 2.914 $&$ 75.968
$&$ 2.193 $&$ -5.890 $ \\
\hline
\end{tabular}
\caption{Parameters of the Fermionic part of the model.}  
\label{table2}
\end{center}
\end{table}

To fix the model parameter the values of the decay constant and meson masses are used  strictly at $T = 0$, $\mu = 0$, and $V = \infty$.

To study the finite volume effects on the thermodynamics of strongly interacting matter the PNJL grand canonical potential in the mean-field approximation in the $SU(3)_{_f}$ sector can be written as
\begin{equation}\label{22}
\begin{split}
  \Omega^{\prime}(\Phi,\bar{\Phi},\sigma_{_f},T,\mu_{_f})=u^{\prime}[\Phi,\bar{\Phi},T]+
  2g_{_s}\sum_{{f=u,d,s}}(\sigma^{2}_{_f}-\frac{g_{_D}}{2}\sigma_{_u}\sigma_{_d}\sigma_{_s})
  \\-T\sum_{_n}\int^{\infty}_{_\lambda}\frac{d^{3}p}{(2\pi)^{3}}Tr ln\frac{S^{-1}(i\omega_{_n},\vec{p})}{T}
\end{split}
\end{equation}
where $\omega_{_n}=\pi T(2n+1)$ are Matsubara frequencies for fermions. In momentum space, the inverse quark propagator is given by
\begin{equation}\label{23}
  S^{-1}=\gamma_{_0}(p^{0}+\hat{\mu}-iA_{_4})-\vec{\gamma}.\vec{p}-\hat{M}.
\end{equation}
Using the identify Tr ln(X) = ln det(X), we get
\begin{equation}\label{24}
\begin{split}
\Omega^{\prime}=u^{\prime}[\Phi,\bar{\Phi},T]+
2g_{_s}\sum_{{f=u,d,s}}(\sigma^{2}_{_f}-\frac{g_{_D}}{2}\sigma_{_u}\sigma_{_d}\sigma_{_s})
-6\sum_{_f}\int^{\Lambda}_{_\lambda}\frac{d^{3}p}{(2\pi)^{3}}E_{_{p{_f}}}\Theta(\Lambda-|\vec{p}|)-2\\
\sum_{_f}T\int^{\infty}_{_\lambda}\frac{d^{3}p}{(2\pi)^{3}}ln[1+3(\Phi+\bar{\Phi} exp(\frac{-(E_{_{p{_f}}}-\mu_{_f})}{T}))
 exp(\frac{-(E_{_{p{_f}}}-\mu_{_f})}{T})+exp(\frac{-3(E_{_{p{_f}}}-\mu_{_f})}{T})]-2\\
\sum_{_f}T\int^{\infty}_{_\lambda}\frac{d^{3}p}{(2\pi)^{3}}ln[1+3(\bar{\Phi}+\Phi exp(\frac{-(E_{_{p{_f}}}+\mu_{_f})}{T}))
 exp(\frac{-(E_{_{p{_f}}}+\mu_{_f})}{T})+exp(\frac{-3(E_{_{p{_f}}}+\mu_{_f})}{T})]\\
\end{split}
\end{equation}

\begin{equation}\label{25}
\Omega^{\prime}=\Omega-\kappa T^{4}ln \jmath[\Phi,\bar{\Phi}]
\end{equation}
where $ E_{_{p{_f}}}=\sqrt{(p^{2}+M^{2}_{_f})}$ is the single quasi-particle energy and $\lambda = \frac{\pi}{R}$, R being the lateral size of the finite volume system. In the above integrals, the medium-dependent integrals have been extended to infinity, whereas the vacuum integral has a cut-off $\Lambda$.
The lower cut-off of $\lambda$ has been set as the lower limit for both the vacuum integrals and medium-dependent integrals.\\

\maketitle
\section{\label{sec:level1} Results
}
In this present paper, we have discussed the  higher-order moments ($M$, $\sigma$, $S$, $\kappa$) of net-baryon and their volume independent products ($ s\sigma, \kappa\sigma^{2}$) in the 3 flavors finite volume finite density Polyakov loop enhanced Nambu-Jona-Lasinio (PNJL) model. In heavy ion collisions, these moments and the susceptibility of the system strongly depend on correlation length $\xi$ produced in the created medium.  The signature of the critical point produce in the medium of strongly interacting matter also has been discussed. Near the critical point, the correlation length diverges, therefore, the susceptibility and the moments. Hence, the non-monotonic behaviour of these moments can be treated as the signature of the critical point. As these moments are system volume dependent, moment products or cumulant ratios can be treated as volume independent of the system. Using the PNJL model, data sets for various cumulants $(C_{1}, C_{2}, C_{3}, C_{4})$ have been obtained at a fixed quark chemical potential for different energies similar to the RHIC Beam Energy Scan (BES) energies (7.7, 11.5, 14.5, 19.6, 27.0, 39.0, 62.4, 130.0 and 200 GeV) and 2.4, 3.0 GeV to compare with net-proton and proton data respectively. In the PNJL model, two sets of finite volume systems have been taken to compare with the infinite volume system. The size of the fireball created in the heavy-ion collision experiments has a finite volume, so finite volume system with the radius R=2fm and R=4fm can be considered for better comparison with the experimental results. The Lagrangian of the NJL model with three flavors has interaction terms with a combination of four quarks and six quarks. Also later eight quarks interaction terms were introduced for vacuum stabilization. Therefore, there are four parameter sets with the combination of 2fm6q, 2fm8q, 4fm6q, and 4fm8q. 
In heavy ion experiments, the correlation length of the colliding system in a strongly interacting medium is of the order of 1.5-3 fm. Therefore, it is essential that the study of the observable should be done within the range and outside of the range in order to study the differences in the magnitude and characteristics of the fluctuations. To determine the temperature (T) and the baryon chemical potential ($\mu_{B}$) of the freeze-out, we have used the suggested equations
\begin{equation}\label{26}
\begin{split}
T(\mu_{B})=a-b(\mu_{B})^{2}-c(\mu_{B})^{4}\\
\mu_{B}(\sqrt{s})=\frac{d}{1+e\sqrt{s}}
\end{split}
\end{equation}
where $a=0.166\pm0.002\: GeV$, $b=0.139\pm0.016\: GeV^{-1}$ and $c=0.053\pm 0.021\:GeV^{-3}$, $d=1.308\pm 0.028\:GeV$, $e=0.273\pm 0.008\:GeV^{-1}$ \cite{38}. Various cumulants $(C_{1}, C_{2}, C_{3}, C_{4})$ which are calculated in the PNJL model with respect to different energies (in GeV unit), are shown in Fig.~1. 

\begin{figure*}[htb]
  \centering
   [$\mathsf{(a)}$]{{\includegraphics[width=8.2cm]{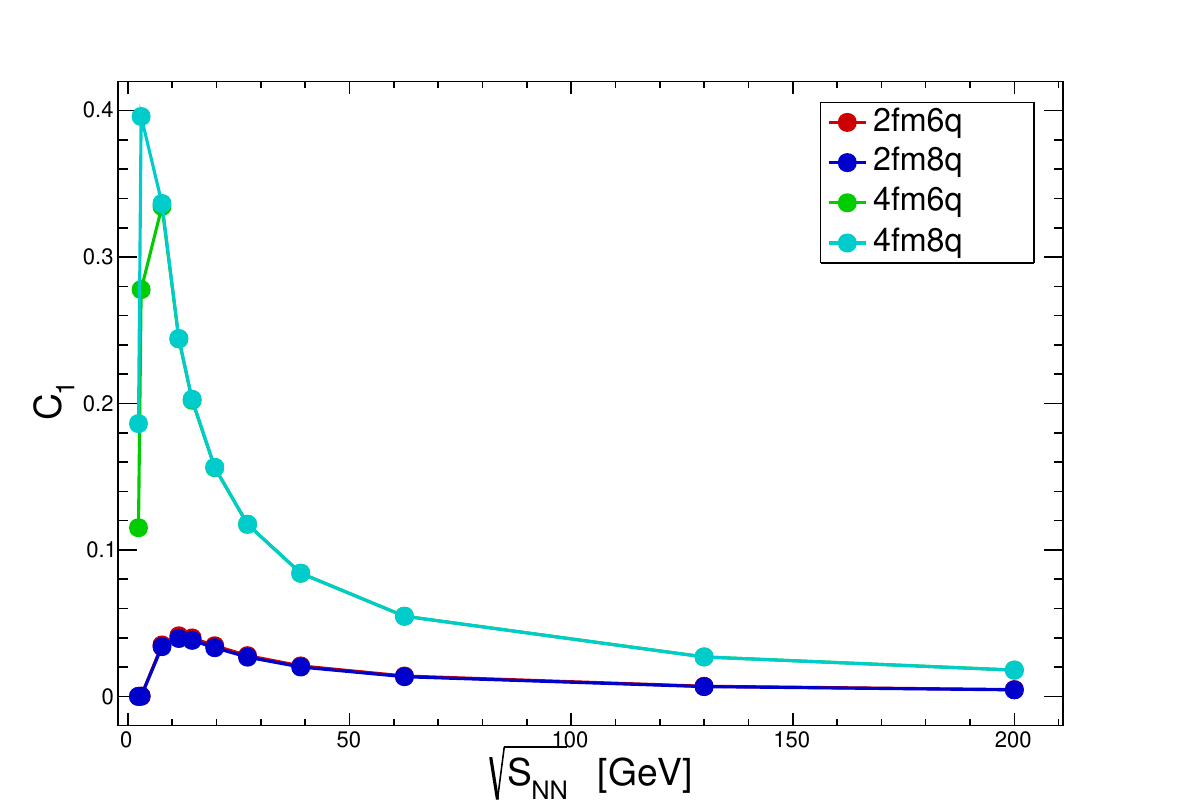} }}
   [$\mathsf{(b)}$]{{\includegraphics[width=8.2cm]{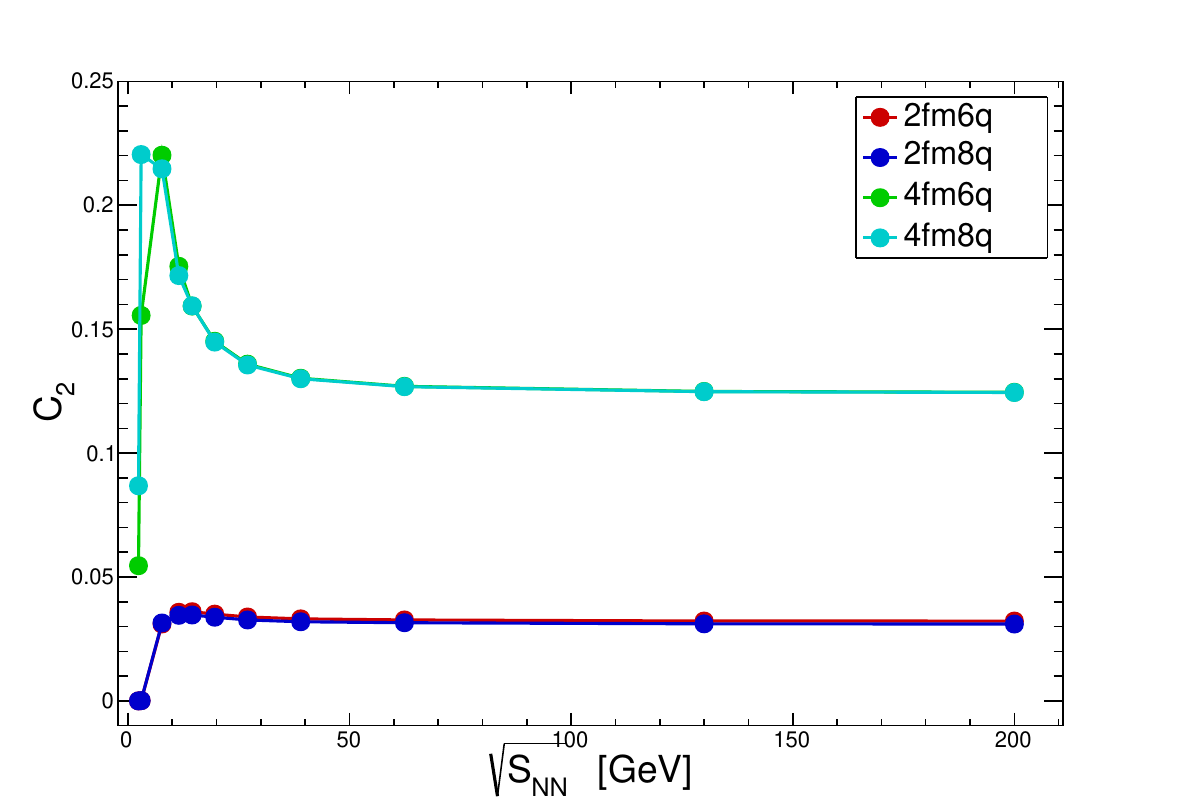} }}
   [$\mathsf{(c)}$]{{\includegraphics[width=8.2cm]{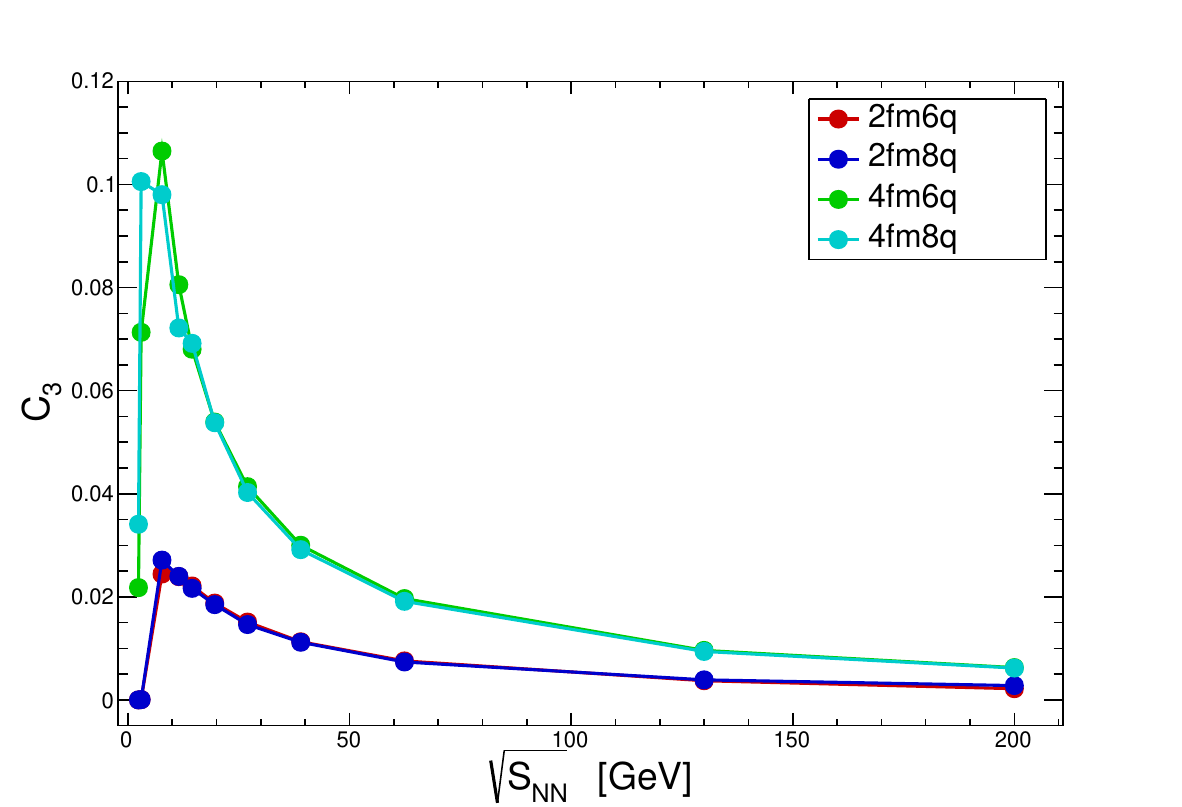} }}
   [$\mathsf{(d)}$]{{\includegraphics[width=8.2cm]{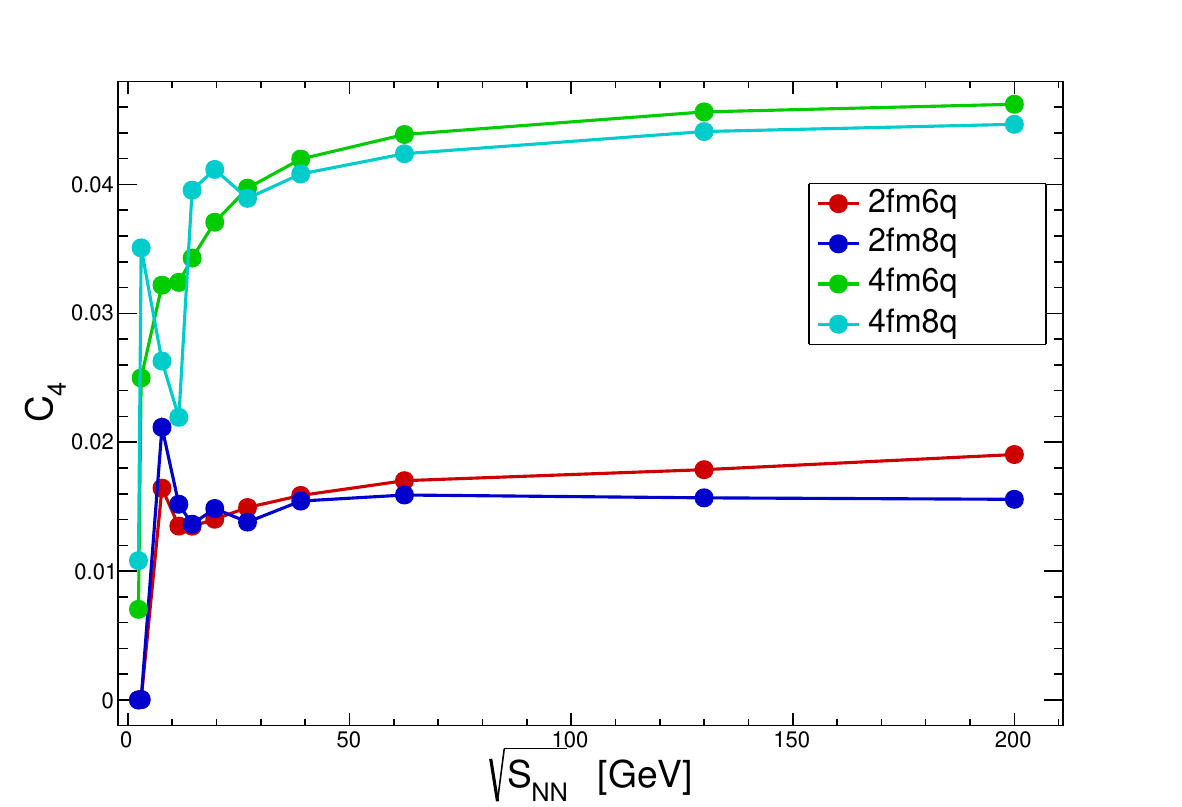} }}
    \caption{(color online) (a) First-order cumulant ($C_{1})$ (b) Second-order cumulant ($C_{2}$)  (c) Third-order cumulant ($C_{3}$) (d) Fourth-order cumulant ($C_{4}$) as a function of energy in PNJL model with 6 quarks and 8 quarks interactions for finite volume systems with R = 2fm and R = 4fm. For system with 6q-PNJL model with R = 2fm is shown in red and R = 4fm is shown in green color. 8q-PNJL with R = 2fm has been shown in blue color and R = 4fm has been shown in cyan color.}
    \label{fig:1}
\end{figure*}

First-order cumulant $(C_{1})$ with respect to energies is shown in Fig.~1(a). In this Figure, 4fm6q (green) and 4fm8q (cyan) curves merge into a single line and decrease exponentially with increasing energy. 2fm6q (red) and 2fm8q (blue) curves also consolidate into a single line and increase firstly and then decrease slowly at lower energies. Towards the higher energies, the gap between the 2fm and 4fm curves diminishes slowly. Fig.~1(b) shows second-order cumulant $(C_{2})$ with respect to different energies. In this Figure, 4fm6q and 4fm8q curves merge to a single line and decrease at a lower energy regime. Towards the higher energies, 2fm and 4fm both curves are independent of energy. Third-order cumulant $(C_{3})$ with respect to different energies are shown in Fig.~1(c). At lower energies, 4fm6q and 4fm8q curves are slightly separated from each other and then merge to a single line. This line decreases exponentially with energy. 2fm6q and 2fm8q curves also merge to a single line and decrease slowly at a lower energy regime. Towards the higher energies, third-order cumulant has the same nature as first-order cumulant. For these cumulants $(C_{1}, C_{2}, C_{3})$, any major fluctuations can not be observed with respect to energies. Fig.~1(d) shows fourth-order cumulant $(C_{4})$ with respect to different energies. In this Figure for 2fm6q (red) data, a dip occurs between 2.4 GeV to 19.6 GeV energies. After that, this curve increases slowly with increasing energy. 2fm8q (blue) line shows similar fluctuations at lower energies but beyond 62.4 GeV energy this curve is independent of energy. In the case of 4fm8q (cyan) data, a dip occurs between 5 GeV to 19.6 GeV energies. Towards the higher energies, the 4fm6q and 4fm8q curves are slowly uprising.

The ratio of first-order cumulant to second-order cumulant $\frac{C_{1}}{C_{2}}$ ($\frac{M}{\sigma^{2}}$) with respect to different energies (left) and the ratio of third-order to second-order cumulant $\frac{C_{3}}{C_{2}}$ or moment product ($S\sigma$) with respect to different energies (right) in PNJL model is shown in Fig.~2. If the energy increases, the ratio $\frac{C_{1}}{C_{2}}$  decreases exponentially but the decreasing value for 4fm data is greater than 2fm data. At lower energies, the 2fm and 4fm curves with different quark numbers (6q and 8q) are slightly separated from each other. But at higher energies, the ratio for four data (2fm6q, 2fm8q, 4fm6q, 4fm8q) is almost the same i.e all lines in the graph are merged into a single line. If the energy decreases, there is some difference between 2fm and 4fm curves with different quark numbers (6q and 8q). The 2fm and 4fm curves with different quark numbers (6q and 8q) are widely separated from each other. For increasing beam energy, the value of $S\sigma$ decreases quantitatively, i.e, the gap between 2fm and 4fm curves with different quark numbers (6q and 8q) slowly diminishes. HRG, STAR, and also UrQMD data for moment products $S\sigma$ and $\kappa\sigma^{2}$ have been collected from Ref. [19, 35, 36] and also added these data in the graphs with respect to different energies which are shown in Fig.~2. The comparison shows that in the case of 2fm data, there is a sameness with the hadron resonance gas model (HRG) structure and 4fm data has the same tendency with STAR data with different values. Any major fluctuations can not be observed in this graph.

\begin{figure*}[htb]
  \centering
   {{\includegraphics[width=8.4cm]{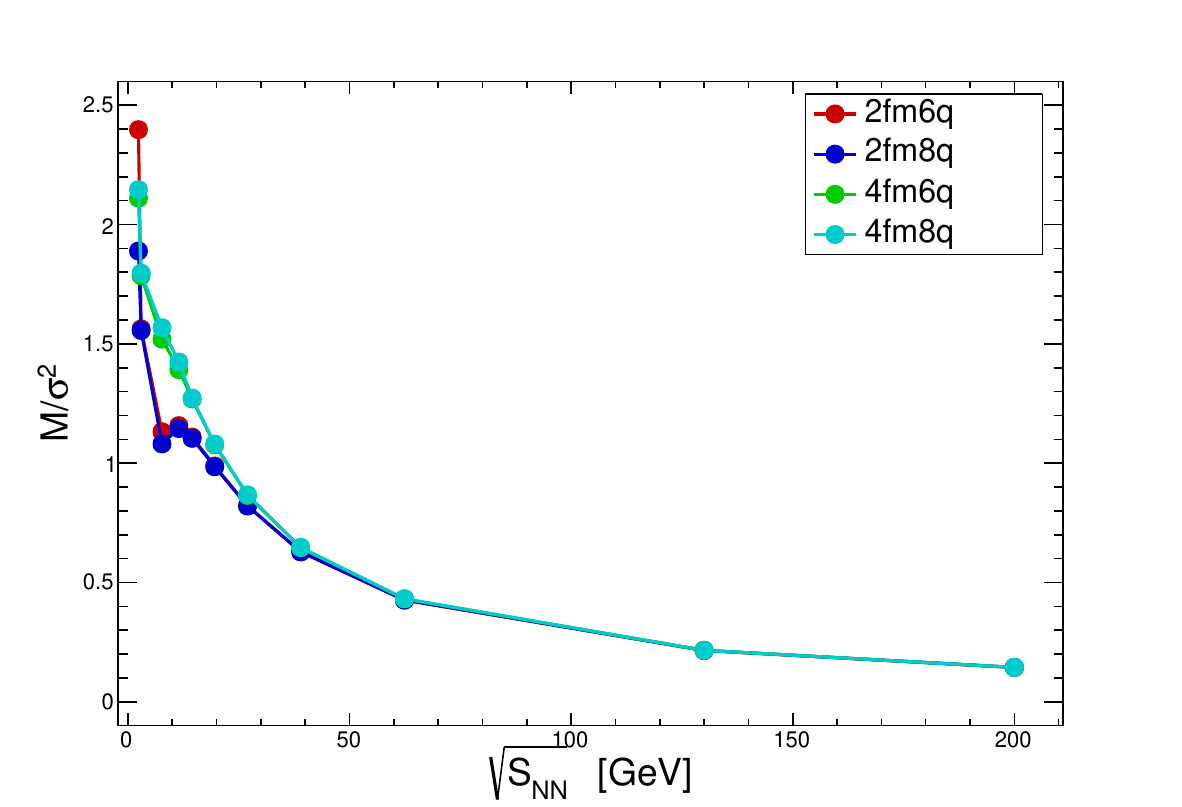} }}
   {{\includegraphics[width=8.4cm]{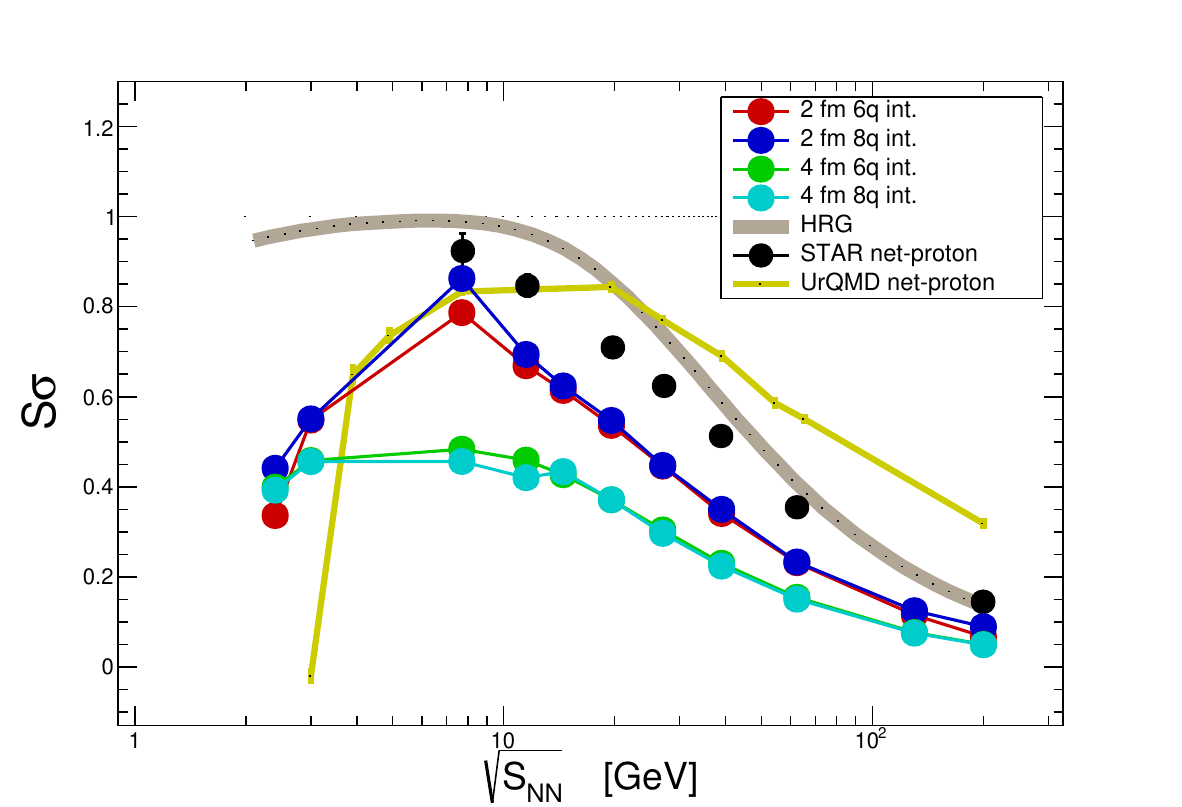} }}
    \caption{(Color online) Left: Ratio of first-order to second-order moments $M/\sigma^{2}$ ($\frac{C_{1}}{C_{2}}$) as a function of the center of mass energy. Right: Ratio of third-order to second-order moment $S\sigma$ ($\frac{C_{3}}{C_{2}}$) as a function of the center of mass energy in PNJL model with 6 quarks and 8 quarks interactions for finite volume systems with R = 2fm and R = 4fm. For system with 6q-PNJL model with R = 2fm is shown in red and R = 4fm is shown in green color. 8q-PNJL with R = 2fm has been shown in blue color and R = 4fm has been shown in cyan color. $S\sigma$ ($\frac{C_{3}}{C_{2}}$) as a function of the center of mass energy are compared with the net-proton data of STAR experiments in black. The results are compared with HRG and UrQMD model calculations in grey and yellow respectively.}
    \label{fig:2}
\end{figure*}

Fig.~3 shows the product of the moment $\kappa\sigma^{2}$ (i.e the ratio of the fourth-order cumulant to the second-order cumulant $\frac{C_{4}}{C_{2}}$) with respect to different energies.  

\begin{figure*}[htb]
  \centering
   {{\includegraphics[width=8.6cm]{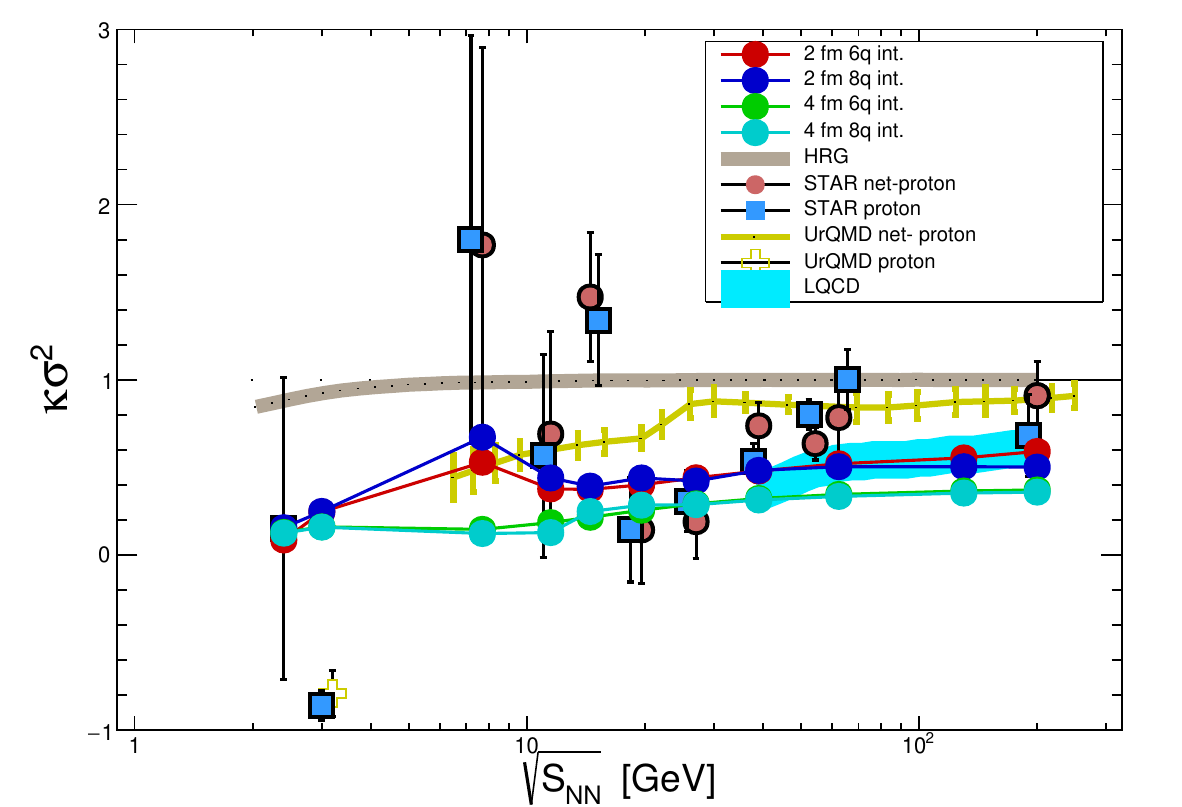} }}
    \caption{(Color online) The Ratio of fourth-order to second-order moment $\kappa\sigma^{2}$ ($\frac{C_{4}}{C_{2}}$) of baryon fluctuations as a function of different center of mass energies in PNJL model with 6 quarks and 8 quarks interactions for finite volume systems with R = 2fm and R = 4fm. For system with 6q-PNJL model with R = 2fm is shown in red and R = 4fm is shown in green color. 8q-PNJL with R = 2fm has been shown in blue color and R = 4fm has been shown in cyan color. The results are compared with the net-proton and proton data of STAR experiments in rust with black circle and azure with black square. The results are compared with HRG and UrQMD model calculations in grey and yellow respectively. Lattice QCD calculation of net-baryon fluctuations from 39 - 200 GeV (in aqua) are also compared. The LQCD (39-200 GeV) data is taken from Ref. \cite{39}.}
    \label{fig:3}
\end{figure*}

These results are compared with the STAR experimental result of net-proton higher order moments calculation from RHIC BES-I data (rust with black circle border) which can serve as the proxy of the net-baryon study of the same having the same dependency as star proton (azure with black square border) measurements are also compared. Calculations are also compared with HRG model calculation (grey line) and UrQMD model calculations (yellow line) from 2 to 200 GeV. Lattice QCD calculations (aqua band) from 39 to 200 GeV are also compared. Below 60 GeV energy, current results show deviation from being constant. For 2fm6q and 2fm8q calculations, fluctuations below 15 GeV are seen. Whereas in 4fm study slowly increasing trend is seen and no fluctuations are observed. Towards the higher energy, these curves show almost independent behaviour with respect to the energy. For decreasing the system size from 4fm to 2fm, a dip kind of structure occurs between 2.4 GeV to 20 GeV energy, after that this curve increases slowly for 2fm and 4fm and becomes independent of energy above 60 GeV. The fluctuation shown in this study has a similar trend to the experimental measurement from STAR at RHIC. The  STAR data is confirming this agreement as STAR data passes through similar fluctuations near the same region of concern. The 4fm data has lower values compared to 2fm data, the HRG, and UrQMD model calculations. The present study matches with Lattice QCD calculations from 39 to 200 GeV and shows a similar tendency. 

\begin{figure*}[htb]
  \centering
   {{\includegraphics[width=8.6cm]{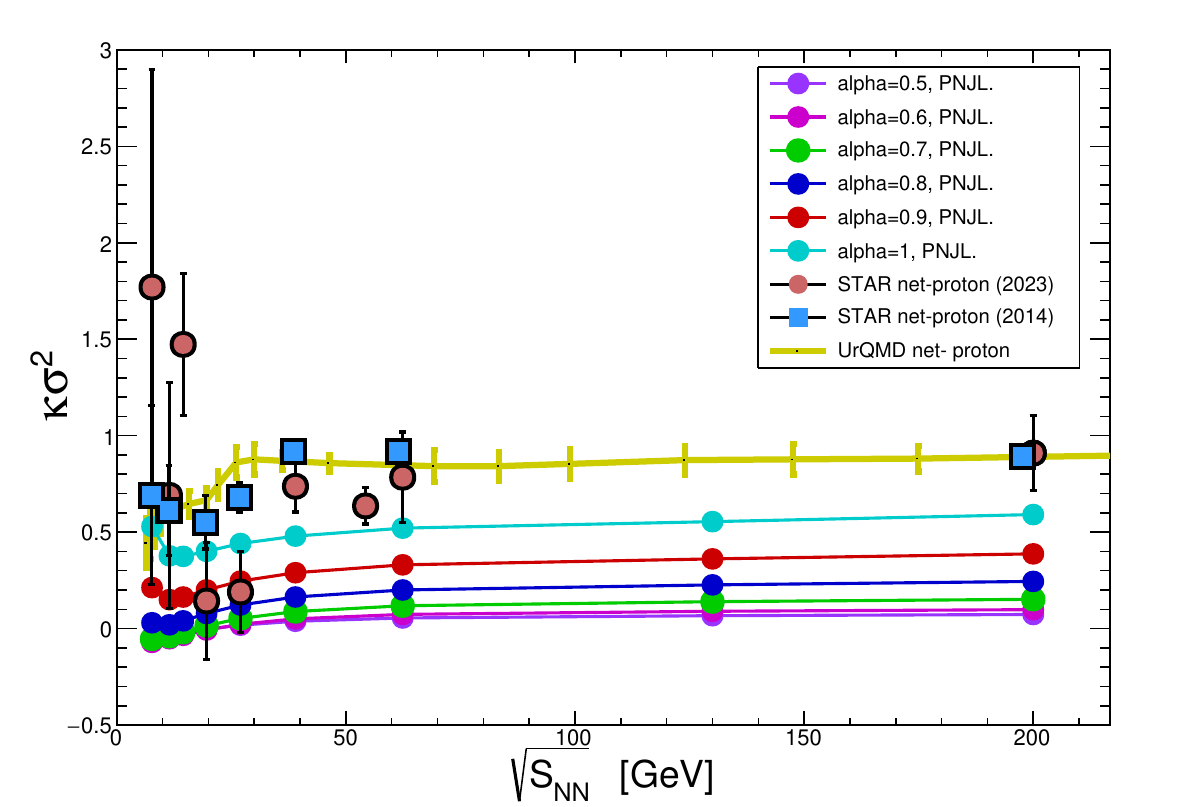} }}
    \caption{(Color online) The Ratio of fourth-order to second-order moment $\kappa\sigma^{2}$ ($\frac{C_{4}}{C_{2}}$) for fixed acceptance fraction alpha $(\alpha)$ as a function of different center of mass energies in PNJL model with 6 quarks interactions for finite volume systems with R = 2fm. The results are compared with the net-proton of STAR experiments in rust with black circle (2023) and azure with black square (2014). The results are compared with UrQMD model calculations in yellow respectively.}
    \label{fig:3}
\end{figure*}

Fig.~4 shows $\kappa\sigma^{2}$ as a function of energy for fixed acceptance fraction $(\alpha)$ for  system with 6q-PNJL model with R = 2fm. The result for $\alpha$ - dependence of $\kappa\sigma^{2}$ is calculated by using Eq. (1). The graph displays the value of $\kappa\sigma^{2}$ within the subvolume system that is obtained from total net baryon number at a particular energy, with each data point representing the acceptance percent. These results are compared with the STAR experimental result of net-proton higher order moments calculation from RHIC BES-I data (rust with black circle border) (2023) and (azure with black square border) (2014) which can serve as the proxy of the net-baryon. At lower energy regimes, the $\alpha$ dependent results exhibit small fluctuations; however, as energies increase, the results remain energy-independent. This figure also demonstrates that the fluctuation at lower energy regions diminishes as the value of alpha ($\alpha$) decreases.

\maketitle
\section{\label{sec:level1}Summary and Conclusions
}
We have studied higher-order moments and moment products of net-baryon in the PNJL model at energies similar to RHIC beam energy scan (BES) energies. These higher order moments like the skewness and the kurtosis are considered to be the observables for CEP. The fluctuations of net baryon higher order moments and their volume independent moment products (cumulant ratios) in the nuclear matter using the PNJL model have been discussed. The various correlations created in the heavy ion collision experiments have been discussed. After fitting the pressure in Taylor series expansion these correlations have been extracted around the finite charge, baryon and strangeness chemical potentials that are obtained from the freeze-out curve. The order parameter ($\phi$) of the Polyakov loop in the PNJL model confirms the existence of the critical point and the critical region. The results are shown for PNJL model for two finite volume systems with lateral size of R = 2fm and R = 4fm with 6 quark and 8 quark interactions. The results of volume independent moment products (cumulant ratios) have been compare with available STAR net proton data, STAR proton data, Lattice QCD calculations, theoretical HRG and UrQMD model. This study concludes that in the case of 4fm data, there is a likeness between HRG, UrQMD structures at energies above 30 GeV.  In the case of 2fm data, $\kappa\sigma^2$ value matched with LQCD calculations above 39 GeV. Fluctuation is observed at lower energies which may be because of the influence of the critical region. STAR net-proton and proton data having similar fluctuation at lower beam energies are confirming this inference. The deviation of the observable from the theoretical values may be due to the influence of the critical region around that energy. Due to the fact that the order of magnitude of the deviations for both the critical region and critical point is unknown, it is very difficult to predict whether the freeze-out went through the critical region or the critical point. STAR data may also pass through this critical region and may need more intermediate data points with higher statistics to confirm the exact location of the critical point. This study confirms that the study of higher order moments of the conserved quantities is a good tool to locate the QCD critical point. In the presence of a critical point (CP) these higher-order moments would show non-monotonic behavior as a function of energy.  An increase in the value of fluctuation at lower energy rather than higher energy signifies that a critical region may exist in a lower energy regime, which is also been found in the experimental results. This current study has a similar dependency as a function of energy with STAR measurement of net-proton multiplicity and implies that the QCD critical region may exist at RHIC lower energies below 19.6 GeV. An enhanced fluctuation at low collision energy might indicate the location of the critical. We have studied how the higher-order moment product ($\kappa\sigma^{2}$) of the net baryon varies with system acceptance which can serve as a proxy for the same. The net proton distribution measured in experiments, depends on the value of the parameter alpha $(\alpha)$ describing the subsystem where the fluctuations are measured. If the acceptance is sufficiently large to achieve the thermodynamic limit and therefore capture all relevant physics, the subensemble acceptance method can be used for an arbitrary equation of state. This method makes it possible to compare experimental data on conserved charge higher order moments directly with grand-canonical susceptibilities calculated theoretically within effective QCD theories and lattice QCD simulations. These $\alpha$ dependence findings are very beneficial for the current experimental endeavor to investigate the QCD phase structure using fluctuation measurements.




\begin{acknowledgments}
P.Deb would like to thank Women Scientist Scheme A (WOS-A) of the Department of Science and Technology (DST) funding with grant no SR/WOS-A/PM-10/2019 (GEN). The authors would also like to thank Ankur Majumdar (RKMRC) for the valuable discussion. Part of the work has been presented in the DAE-BRNS High Energy Physics Symposium 2022.
\end{acknowledgments}


\end{document}